\documentclass[fleqn,usenatbib]{mnras}

\usepackage{newtxtext,newtxmath}

\usepackage[T1]{fontenc}
\usepackage{ae,aecompl}

\usepackage{graphicx}	
\usepackage{amsmath}	
\usepackage{amssymb}	
\usepackage{natbib}
\usepackage{gensymb}
\usepackage{ulem}

\title[Producing Synthetic Polarized Dust Maps using VChG]{Producing Synthetic Maps of Dust Polarization using Velocity Channel Gradient Technique}

\author[Z. Lu, A. Lazarian \& D. Pogosyan]{
Zekun Lu,$^{1}$\thanks{E-mail: zekun2@ualberta.ca}
A. Lazarian$^{2}$ and Dmitri Pogosyan$^{1}$ 
\\
$^{1}$Department of Physics, University of Alberta, Edmonton, T6G 2E1, Canada \\
$^{2}$Department of Astronomy, University of Wisconsin-Madison, Madison, WI 53706-1507, USA
}

\date{Accepted XXX. Received YYY; in original form ZZZ}

\pubyear{2019}

\hypersetup{final}
\begin{document}
\label{firstpage}
\pagerange{\pageref{firstpage}--\pageref{lastpage}}
\maketitle

\begin{abstract}
In modern cosmology, many efforts have been put to detect primordial B-mode of cosmic microwave background (CMB) polarization from the gravitational waves generated during inflation. Considering the foreground dust contamination of microwave polarization maps, it is essential to obtain a precise prediction for polarization in dust emission. In this work, we show a new method to produce synthetic maps of dust polarization in magnetized turbulent ISM from more abundant high-resolution HI data. By using Velocity Channel Gradient (VChG) technique, we are able to predict both direction and degree of dust polarization by investigating spectroscopic HI information in position-position-velocity (PPV) space. We applied our approach to The Galactic Arecibo L-band feed Array HI (GALFA-HI) data, and find a good correspondence between synthesized maps and PLANCK's polarization measurements at 353 GHz.
\end{abstract}




\section{Introduction}
\label{sec:introduction}
The galactic polarized dust emission is of interest for our understanding of the interstellar medium (ISM; \citealt{Heiles2000AJ....119..923H}; \citealt{Fissel2016ApJ...824..134F}) and, as an important foreground for cosmology, essential for the detection of the inflationary gravitational wave (IGW) B-mode polarization in the cosmic microwave background (CMB; see \citealt{Lazarian&Finkbeiner2003NewAR..47.1107L} for a review). The dust will absorb the interstellar radiation and re-radiate it as partially polarized light, which would contaminate the polarization from the CMB. For example, the pioneering claim of detection of the primordial B modes by BICEP2 Collaboration \citep{BICEP2PhysRevLett.112.241101} could be interpreted as the contribution of Galactic dust emission (\citealt{Raphaeletal20141475-7516-2014-08-039}; \citealt{PAR.Adeetal.2015PhysRevLett.114.101301}). Therefore, a precise measurement of dust polarization is needed before we could obtain a conclusive detection of IGW B modes. 

\textit{Planck} Collaboration (\citealt{Planck_Collaboration2015A&A...576A.104P},  \citeyear{Planck_Collaboration_2016_Planck2015results.I2016A&A...594A...1P}) released the full-sky polarization map at 353 GHz, a frequency dominated by thermal dust emission, which could work as the foreground subtraction of the CMB. However, the \textit{Planck} maps are noise dominating at high Galactic latitudes. Thus, it is important to look for alternative methods to produce polarization map.

It was noticed in \cite{Clark2015PhRvL.115x1302C} that HI emission filaments are aligned with interstellar magnetic field, which was proposed as an independent method to trace magnetic fields. Later in (\citealt{Clark2018ApJ...857L..10C}), such method was developed to predict degree of polarization.

In parallel to this direction suggested by S. Clark and her collaborators on the basis of observational studies, a theory-based approach was developed by the Lazarian's research group (see \citealt{GL2017ApJ...835...41G}, \citealt{YL2017-HI-ApJ...837L..24Y}, \citealt{LY2018-Channel-ApJ...853...96L}, \citealt{Lazarian2018-MA-ApJ...865...46L}, \citealt{Hu2018MNRAS.480.1333H}).  This approach of tracing interstellar magnetic field is based on the property of velocity gradients to be aligned perpendicular to magnetic field in MHD turbulence. Among the modifications of the technique, the  Velocity Channel Gradients (VChGs) technique described in \cite{LY2018-Channel-ApJ...853...96L} was shown the most promising. The technique makes use of the theory of non-linear mapping of turbulent motions from the Position-Position-Position (PPP) space into the Position-Position-Velocity (PPV) space developed in \citeauthor{LP2000ApJ...537..720L} (\citeyear{LP2000ApJ...537..720L}, \citeyear{LP2004ApJ...616..943L}). 

While the technique proposed by \cite{Clark2015PhRvL.115x1302C} was from the very beginning aimed at obtaining the information that can be used to study foreground polarization, the VChGs were used mostly for magnetic field studies, in particular, for magnetic field studies in molecular clouds \citep{Hu2019NatAs.tmp..334H}. The studies of magnetic field in HI were also performed, but they were aimed at studies of magnetic fields and magnetization of media (\citealt{LY2018-Channel-ApJ...853...96L}, \citealt{Lazarian2018-MA-ApJ...865...46L}) rather than on foreground studies. In this paper we explore the latter application of the VChGs.

In spite of the progress that we have made, as a pioneering method, the VChG can be improved from several aspects. For one thing, in the original idea of VChG, the description of polarization is incomplete, which only includes the direction but lack the degree of polarization. For another, as for the statistical distribution of local gradient angles, the previous method gives a phenomenological fitting but lacks a theoretical/mathematical understanding.
In this paper, firstly, we improve the VChG method by constructing pseudo Stokes parameters, which makes it possible to estimate degree of polarization. Secondly, we analytically study the local statistical property of gradient angle, which helps us further understand the behaviour of VChG.

This paper unfolds as follows. Following the introductory part in \autoref{sec:introduction}, we explain the theoretical bases of our method in \autoref{sec:theory} and introduce our method in \autoref{sec:Method}. In \autoref{sec:numeric}, we numerically test our method and then apply it into observation in \autoref{sec:observation}. Finally, we discuss the limitation and future plan of our method in \autoref{sec:discussion} and briefly summarize the paper in \autoref{sec:summary}.

\section{Theoretical considerations}
\label{sec:theory}
\subsection{MHD turbulence}
The foundations of the modern MHD turbulence theory were presented in \cite{GS1995ApJ...438..763G} (see also a book by Beresnyak \& Lazarian 2019). For the velocity gradient studies as well as for the sister technique that uses synchrotron intensity and polarization gradients (see \citealt{Lazarian2017ApJ-Synchrotron...842...30L}, \citealt{LY2018bApJ...865...59L}) it is important that the gradients of velocities and magnetic field in MHD turbulence are aligned in respect to the {\it local} direction of magnetic field rather than the mean magnetic field direction.\footnote{Incidentally, the original formulation of the Goldreich-Sridhar theory was implicitly using the mean magnetic field.} The {\it local} direction is the direction of the magnetic field of surrounding the turbulent eddies. According to the turbulent reconnection theory 
 (\citealt{LV1999ApJ...517..700L}) the magnetic field does not constrain the motions of the eddies that mix magnetized matter if the velocities of the eddies are perpendicular to the magnetic field. Indeed, the theoretical prediction is that the magnetic "knots" arising from this type of motion are resolved within one eddy turnover. As a result, most of the turbulent energy gets in the form of such eddies mixing up magnetic field surrounding the eddies without much bending of the magnetic field.  
 
 The concept of the eddies tracing the {\it local} magnetic field was confirmed in numerical simulations (see \citealt{Cho&Vishniac2000ApJ...539..273C}; \citealt{Maron&G2001ApJ...554.1175M}; \citealt{Cho&L&V2002ApJ...564..291C}) and it has become an essential element of the modern theory of MHD turbulence. This concept is also crucial for understanding how magnetic field tracing with velocity gradients works. 
 
 In this {\it local} magnetic field frame, the relation between parallel and perpendicular scales of the eddies is given by \cite{GS1995ApJ...438..763G} as follows
\begin{equation}
l_\parallel^{-1}V_A\approx l_\perp^{-1}u_l ~,
\end{equation}
where $V_A$ is the Alfv\'en speed, $u_l$ is the eddy velocity, and $l_\parallel$ and $l_\perp$ are the eddy scales parallel and perpendicular to the local direction of the magnetic field, respectively. The elongated eddies have the largest velocity gradient, which is perpendicular to the longest axes of eddies. Thus, we expect the direction of the maximum velocity gradient to be perpendicular to the {\it local} magnetic field (aforementioned elongated direction). In this way, taking such 90 degrees angle difference, the velocity gradients can trace the directions of the {\it local} magnetic field. 

The most accepted way of magnetic field tracing is using dust polarization. According to dust alignment theory, the long axes of dust grains are aligned perpendicular to {\it local} magnetic fields (see \citealt{Lazarian2007JQSRT.106..225L} for a review). The Planck measurements are affected by the polarization arising from the emission arising from aligned dust. This polarization is along the long dust grain axes and, as a result, dust emission polarization and velocity gradients are both perpendicular to magnetic field. This opens a possibility to use velocity gradients to predict the polarization arising from the aligned dust and interfering with the CMB polarization studies. Indeed, atomic hydrogen and dust are well mixed at high galactic latitudes. Therefore the velocity gradients obtained with 21 cm emission can be used to predict the polarization from the dust. In this paper, we explore the accuracy of this approach.

\subsection{Gradient methods to study magnetic field}

Let us briefly discuss the mathematical foundations of the gradient methods in application to study of the direction of the magnetic field. Observing emission from the turbulent media, one constructs the sky maps of different observables that describe the emission. First of all, this is the intensity of the emission in PPV (position-position $\mathbf{X}$-velocity $v$) space
$I(\mathbf{X},v)$, and the related integrated quantities, such as the total intensity  and velocity centroids which for optically thin lines are
$I_c(\mathbf{X}) \propto \int dv I(\mathbf{X},v)$ and
$VC(\mathbf{X}) \propto \int v dv I(\mathbf{X},v) $
respectively. The maps represent random fluctuating fields.

The simplest local statistical measure of the gradient of a random field 
$f(\mathbf{X})$ is the gradient covariance tensor
\begin{equation}
\label{eq:gradcov}
\sigma_{\nabla_i \nabla_j} \equiv \left\langle \nabla_i f(\mathbf{X}) \nabla_j f(\mathbf{X}) \right\rangle
= \nabla_i \nabla_j D(\mathbf{R}) | _{R\to 0} ~,
\end{equation}
which is the zero separation limit of the second derivatives
of the field structure function 
$D(\mathbf{R})\equiv \left\langle\left( f(\mathbf{X+R}) - f(\mathbf{X})\right)^2\right\rangle$.

For a statistically isotropic field, the covariance
of the gradients is isotropic, $\sigma_{\nabla_i \nabla_j} = \frac{1}{2} \delta_{ij} \Delta D(R) | _{R\to 0}$. However,
as was studied in \cite{LP2012ApJ...747....5L} for synchrotron, \cite{Kandel2016MNRAS.461.1227K} for velocity channel intensities and \cite{Kandel2017MNRAS.464.3617K-velocity-centroid} for velocity centroids, in the presence of the magnetic field, the structure function
of the signal becomes orientation dependent, depending
on the angle between $\mathbf{R}$ and the projected direction of the magnetic field. This anisotropy is retained in the limit $\mathbf{R} \to 0$ and results in non-vanishing  traceless part of the gradient covariance tensor
\begin{align}
\sigma_{\nabla_i \nabla_j} &- \frac{1}{2} \sum_{i=1,2} \sigma_{\nabla_i \nabla_i } = 
\nonumber \\
& \frac{1}{2}\left(
\begin{array}{cc}
\left(\nabla_x^2  - \nabla_y^2\right)  D(\mathrm{R})
& 2 \nabla_x \nabla_y  D(\mathrm{R}) \\
2 \nabla_x \nabla_y  D(\mathrm{R}) &
\left(\nabla_y^2 - \nabla_x^2\right)  D(\mathrm{R})\\
\end{array}
\right)_{\mathbf{R} \to 0} 
\ne 0
\end{align}
The eigendirection of the covariance tensor that corresponds to the largest eigenvalue (``the direction of the gradient'') then makes 
an angle $\theta$ with the coordinate x-axis
\begin{equation}
\label{eq:theta_grad}
\tan \theta =
\frac{ 2 \nabla_x \nabla_y D}
{\sqrt{\left(\nabla_x^2 D - \nabla_y^2 D \right)^2 + 4 \left( \nabla_x \nabla_y D \right)^2}+\left( \nabla_x^2 - \nabla_y^2 \right)D}~.
\end{equation}

Anisotropic structure function can be decomposed in angular harmonics.
In Fourier space, where
\begin{equation}
D(\mathbf{R}) = -\int d\mathbf{K} \; P(\mathbf{K}) e^{i \mathbf{K} \cdot \mathbf{R}} ~,
\end{equation}
this decomposition is over the dependence of the power spectrum $P(\mathbf{K})$
on the angle of the 2D wave vector $\mathbf{K}$.
Denoting the coordinate angle of $\mathbf{K}$ by $\theta_K$ and that of the projected magnetic field as $\theta_H$, we have for the spectrum
\begin{equation}
P(\mathbf{K}) = \sum_n P_n(K) e^{i n (\theta_H - \theta_K)}
\end{equation}
and for the derivatives of the structure function
\begin{align}
&\nabla_i \nabla_j D(\mathbf{R}) =  \nonumber\\
&=\sum_n \int K^3 P_n(K)
\int d\theta_K e^{i n (\theta_H - \theta_K)} 
e^{i K R \cos(\theta_R - \theta_K)} \hat K_i \hat K_j ~,
\end{align}
where hat designates unit vectors, namely $\hat K_x = \cos \theta_K$ and $\hat K_y = \sin \theta_K$. Performing integration over
$\theta_K$, we obtain the traceless anisotropic part 
\begin{align}
& (\nabla_x^2 - \nabla_y^2)  D(\mathbf{R}) =  2 \pi \sum_n i^n e^{ i n (\theta-\theta_H)}  \; \times \\
& \times \int dK K^3 J_n(k R) \left( P_{n-2}(K) e^{i 2 \theta_H} + P_{n+2}(K) e^{-i 2 \theta_H} \right)  \nonumber  \\
& \nabla_x \nabla_y D(\mathbf{R}) = \pi \sum_n i^{n+1}
e^{ i n (\theta-\theta_H)}  \; \times \\
& \times \int dK K^3 J_n(k R) \left(-P_{n-2}(K) e^{i 2 \theta_H} + P_{n+2}(K) e^{-i 2 \theta_H} \right)
\nonumber
\end{align}
In the limit $R \to 0$, only $n=0$ term for which $J_0(0)=1$ survives and we have
\begin{align}
(\nabla_x^2 - \nabla_y^2) D(\mathbf{R}) & = 
\left[ 2 \pi  \int dK K^3 P_2(K) \right] \cos 2 \theta_H  \\
2 \nabla_x \nabla_y D(\mathbf{R}) & = \left[
2 \pi \int dK K^3 P_2(K)  \right] \sin 2 \theta_H \end{align}
Notice that anisotropy of the gradient variance is determined by the quadrupole of the power spectrum (and structure function). Substituting this result into
\autoref{eq:theta_grad}, we find that the eigendirection of the gradient variance has the form
\begin{equation}
\tan \theta = \frac{ A \sin 2 \theta_H}{ |A| + A \cos 2 \theta_H} =
\left\{
\begin{array}{rl}
\tan \theta_H & A > 0 \\
-\cot \theta_H & A < 0 
\end{array}
\right.
\end{equation}
and is either parallel or perpendicular to the direction of the magnetic field,
depending on the sign of $A \propto \int dK K^3 P_2(K) $, i.e the sign of the spectral quadrupole $P_2$.

Since the direction of the magnetic field that we aim to track is unsigned,
it is appropriate to describe it as an eigendirection of the rank-2 tensor, rather than a vector.  This naturally leads to the mathematical formalism of Stokes parameters.
As the local estimator of the angle $\theta$ via the gradients, we can introduce pseudo-Stokes parameters 
\begin{align}
\widetilde{Q} & \propto (\nabla_x f)^2 - (\nabla_y f)^2 \propto \cos 2\theta \\
\widetilde{U} & \propto 2 \nabla_x f \nabla_y f  \propto \sin 2 \theta
\end{align}
so that
\begin{equation}
\frac{\widetilde{U}}{\widetilde{Q}}=\tan 2\theta \sim \tan 2 \theta_H
\end{equation}
In the next section, we describe the exact procedure for the
estimator that we use in this paper.

The pseudo Stokes parameters naturally connect the gradient techniques with
polarization studies. More exactly,
both for synchrotron (\citealt{LP2012ApJ...747....5L}; \citealt{Kandel2018MNRAS.478..530K}) and thermal dust emission (\citealt{Clark2015PhRvL.115x1302C}; \citealt{Caldwell2017ApJ...839...91C}; \citealt{Kandel2018MNRAS.478..530K}, see Crutcher 2010 and ref. therein), we expect the true polarization Stokes parameters to be
\begin{align}
Q & \propto \int dz (H_x^2 - H_y^2) \propto \cos 2\theta_H \\
U & \propto \int dz 2 H_x H_y  \propto \sin 2 \theta_H
\end{align}
Thus, the pseudo Stokes parameters constructed from the gradients can be directly
compared with Stokes parameters that probe polarized emission in magnetized medium.

\section{Method} 
\label{sec:Method}
\subsection{Main steps}
Making use of the optically thin emission line maps in PPV space, such as HI 21cm maps, we are able to determine the direction of the magnetic field through our new method. In such method, we focus on obtaining the maximum information from the motions of the gas by using full resolution velocity channel slices of PPV cube. 
It consists of the following steps:
\begin{enumerate}
\item As a first preparatory step, to have robust control of intensity gradient determination at pixel level, full spatial resolution individual velocity channel maps I(\textbf{X},v) are smoothed by a Gaussian filter with $FWHM=3$ pixel. Direction of the gradient
at each pixel $p=(i,j)$ is defined as
\begin{equation}
\label{eq:SimpleGradient}
\theta_p(i,j) = \mathrm{atan2} \left[\frac{I(i,j+1)-I(i,j-1)}{I(i+1,j)- I(i-1,j)}\right] ~,
\end{equation}
which is on the rectangular grid with nodes at coordinates $\mathbf{X}(i,j)$.
Here $(i,j)$ are the indexes of raw pixels. The magnitude of the gradient is not used.
\item The map is then fragmented into large super-pixel blocks, over which the average
direction of the gradient is defined by calculating the mean
$\overline{\cos 2\theta_p}$ and $\overline{\sin 2\theta_p}$ within the block $B=(I,J)$. Here $(I,J)$ are the indexes of super-pixels blocks. Technical details of how averaging is performed are given in section~\ref{sec:finetobigpix}.

Finding as well the total intensity in the block, we define a set of coarse-grained pseudo Stokes parameters for each velocity channel:
\begin{align}
I_B(I,J,v) & = \sum_{p \in B} I_p(v) \\
\widetilde{Q}_B(I,J,v) & = I_B \overline{\cos 2\theta_p}  \\
\widetilde{U}_B(I,J,v) & = I_B \overline{\sin 2\theta_p}
\end{align}
,where tilde signifies the pseudo nature of these ``polarization" parameters (since HI emission is not polarized). The (intensity independent) direction and degree of ``polarization'' in each super-pixel are:
\begin{align}
\theta_B(I,J,v) & = \frac{1}{2} \mathrm{atan}\left[\frac{\overline{\sin 2\theta_p}}{\overline{\cos 2\theta_p}}\right] \\
p_B(I,J,v) &= \sqrt{(\overline{\cos 2\theta_p})^2 + (\overline{\sin 2\theta_p})^2} \le 1
\end{align}
Here we note that averaged $\overline{\cos 2\theta_p}$ and $\overline{\sin 2\theta_p}$ squared do not, in general, add to unity, so the procedure describes depolarization due to variation of directions within the super-pixel block.
\item At the last step of our VChG method,
using additivity property of Stokes parameters when emission is combined, we sum over all velocity channels to obtain the total coarse grained
maps of pseudo Stokes parameters:
\begin{align}
I_{VChG}(I,J) & = \sum_{v} I_B(I,J,v) \\
\tilde{Q}_{VChG}(I,J) & =  \sum_{v} \widetilde{Q}_B(I,J,v) \\
\tilde{U}_{VChG}(I,J) & = \sum_{v} \widetilde{U}_B(I,J,v) 
\end{align}
and of the final direction angle and ``polarization'' degree:
\begin{align}
\theta_{VChG}(I,J) & = \frac{1}{2} \mathrm{atan}\left[\widetilde{U}_B(I,J)/\widetilde{Q}_B(I,J)\right]
\label{eq:VChG_theta} \\
p_{VChG}(I,J) &= \sqrt{\widetilde{Q}_B(I,J)^2 + \widetilde{U}_B(I,J)^2}/I_B(I,J)
\label{eq:VChG_p}
\end{align}
\end{enumerate}

Using Stokes parameters to propagate information about direction of the gradients, as well as the level of their variance within the coarse grained pixel, allows us to use the final $I, \tilde{Q}, \tilde{U}$ maps in two-fold way. As far as the direction only is concerned, they give the prediction for the direction of the magnetic field. But if we also use the observational fact that HI distribution closely follows that of the thermal dust, we can consider our pseudo-Stokes maps as a prediction from HI data for the polarized dust emission.

\subsection{Evaluation of the mean direction on coarse grained map}
\label{sec:finetobigpix}
\begin{figure*}
\includegraphics[width=1.0\textwidth]{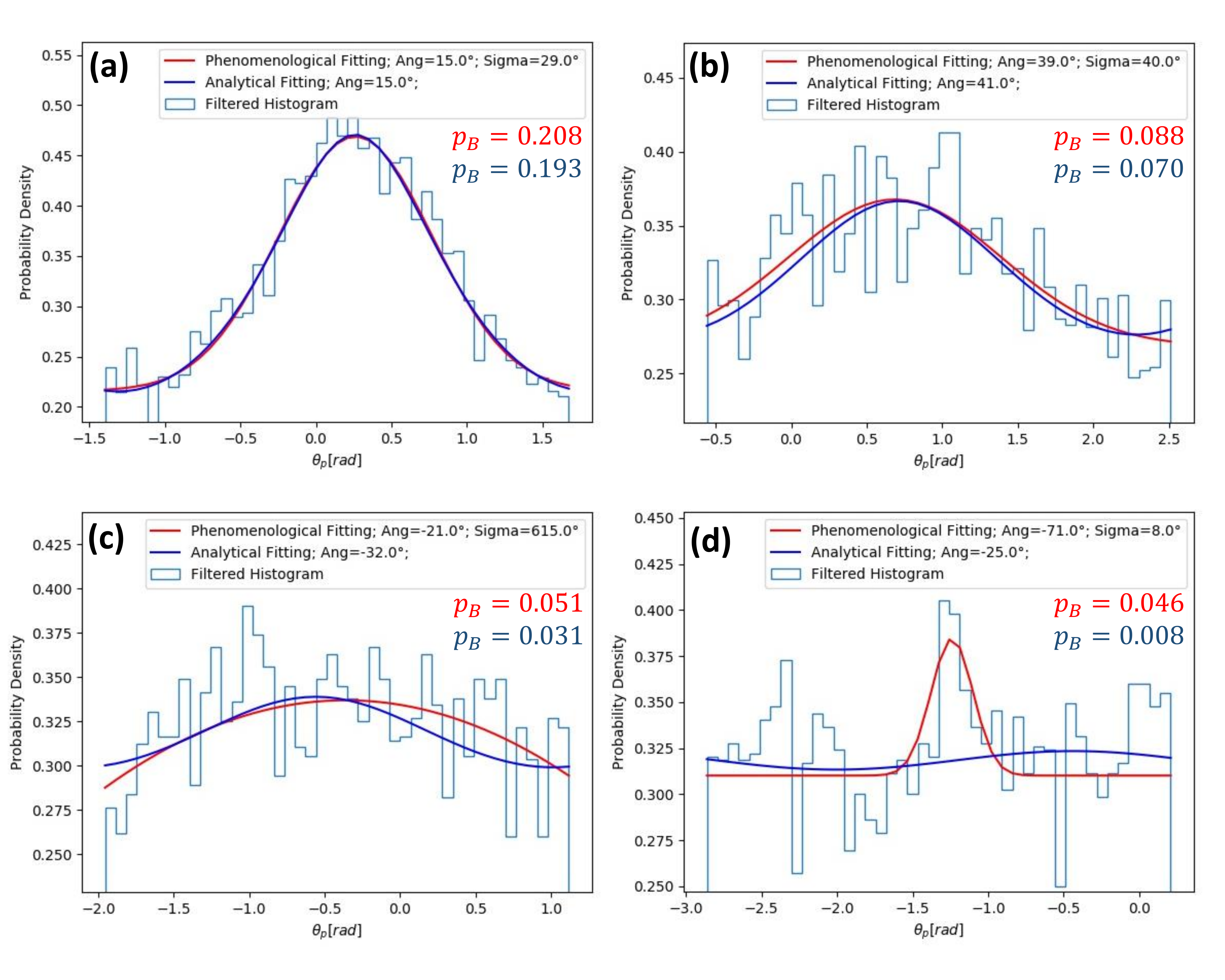}
\caption{Examples of distribution of gradient directions $\theta_p$ within coarse grained pixels. Raw histograms and two model fits are shown. Blue curve - analytical Gaussian distribution for gradients, Equation~\ref{eq:Gaussian-grads}. Red curve - phenomenological fit of Gaussian shape plus constant offset, Equation~\ref{eq:Gaussian-fitting}. Two method's measurements of degree of polarization are also shown in two colors in each panel. Patch coherence and $p_B$ decreases from panel (a) to panel (d). Both models give consistent direction determination for the top row. Panel (c) panel shows the case of wide distribution where non-periodic nature of the Gaussian curve leads to deviations. Panel (d) shows highly noisy pixel where red model pixel some direction above the noise,
while blue model shows practically flat distribution.}
\label{fig:histograms}
\end{figure*}

The fundamental step of our method is to define the averaged direction of PPV maps' gradients within a coarse grained block. We investigate two methods, both based on fitting the measured histogram of angle distribution within coarse grained block with analytical formula.
The difference is the choice of the  fitting model.

First approach has a simple theoretical foundation. As the model distribution of the gradient angle $\theta_p = \mathrm{atan2}\left[\nabla_y I(i,j,v)/\nabla_x (I,i,j,v)\right], ~p \in B$, we take the distribution that follows from assuming the gradients to be Gaussian with the covariance matrix $\widetilde{\sigma}_{ij}$ of Equation~\ref{eq:gradcov}. The expression for this distribution can be easily obtained as:
\begin{align}
\label{eq:Gaussian-grads}
P(\theta_p) &= 
\frac{1}{\pi \sqrt{|\widetilde{\sigma}_{ij}|}} 
\left[
\left(
\begin{array}{c}
\cos\theta_p \\
\sin\theta_p      
\end{array}
\right) 
\widetilde{\sigma}_{ij}^{-1}
\left(
\begin{array}{c}
\cos\theta_p \\
\sin\theta_p      
\end{array}
\right)
\right]^{-1} 
\\
&= \frac{1}{\pi} \times \frac{ 
\sqrt{1 - \widetilde{\mathcal{J}}_2}}
{1- \sqrt{\widetilde{\mathcal{J}}_2} \cos2 \left(\widetilde{\theta}_B-\theta_p\right)}
\end{align}
This distribution function has two parameters: $\widetilde{\mathcal{J}}_2$ and $\widetilde{\theta}_B$. The first is the rotation invariant ratio of the determinant of the traceless part of the covariance matrix and the (half of) trace of the covariance
\begin{equation}
\widetilde{\mathcal{J}}_2 \equiv \frac{\left(\widetilde{\sigma}_{xx}-\widetilde{\sigma}_{yy}\right)^2 + 4 \widetilde{\sigma}_{xy}^2}
{\left(\widetilde{\sigma}_{xx}+\widetilde{\sigma}_{yy}\right)^2}
\end{equation}
and the second is the angle
\begin{equation}
\tan 2 \widetilde{\theta}_B \equiv \frac{2 \widetilde{\sigma}_{xy}}{\widetilde{\sigma}_{xx}-\widetilde{\sigma}_{yy}}
\end{equation}
They can be thought of as estimators (thus tilde) for the coarse grained combinations of covariance components of the gradient.  The distribution in \autoref{eq:Gaussian-grads} is periodic with a period $\pi$ and is normalized to unity on any angular interval of the length of the period, $\int_{\theta_p^*}^{\theta_p^*+\pi} P(\theta_p) d\theta_p = 1$.  Statistically isotropic case
corresponds to $\widetilde{\sigma}_{xx}=\widetilde{\sigma}_{yy}$ and $\widetilde{\sigma}_{xy}=0$, i.e $\mathcal{J}_2=0$ when
Equation~\ref{eq:Gaussian-grads} evaluates to uniform distribution.
Any anisotropy leads to non-zero $\widetilde{\mathcal{J}}_2 > 0$, which on the other hand is bounded
by definition not to exceed unity, $\widetilde{\mathcal{J}}_2 \le 1$.
Note that fitting angular distribution does not determine the trace of gradient covariance
$\widetilde{\mathcal{I}}_1 = \widetilde{\sigma}_{xx}+\widetilde{\sigma}_{yy}$.

The average $\overline{\cos 2 \theta}$ and $\overline{\sin 2 \theta}$  are now readily obtained
\begin{equation}
\left(
\begin{array}{c}
\overline{\cos 2\theta_p}\\
\overline{\sin 2\theta_p} 
\end{array}
\right) = 
\left(
\begin{array}{c}
\cos 2\widetilde{\theta_B}\\
\sin 2\widetilde{\theta_B} 
\end{array}
\right)
\times
\frac{1-\sqrt{1 - \widetilde{\mathcal{J}}_2}}{ \sqrt{\widetilde{\mathcal{J}}_2}}
~. 
\end{equation}
From here, the coarse grained direction is simply given by the angle fit parameter
\begin{equation}
\theta_B = \widetilde{\theta_B} \equiv \frac{1}{2} \mathrm{atan}
\frac{2 \widetilde{\sigma}_{xy}}{ \widetilde{\sigma}_{xx} - \widetilde{\sigma}_{yy}} ~,
\end{equation}
i.e the detected $\theta_B$ is given by the peak position of the fitted
distribution.
The degree of polarization is, as expected, a rotation invariant quantity:
\begin{equation}
p_B = \frac{1-\sqrt{1 - \widetilde{\mathcal{J}}_2}}{\sqrt{\widetilde{\mathcal{J}}_2}} ~.
\end{equation}
It varies from zero for isotropic case $\widetilde{\mathcal{J}}_2=0$
to unity for maximum anisotropy at $\widetilde{\mathcal{J}}_2=1$.

This formalism allows to analyze the effect of noise on our estimators of
the direction and the degree of polarization.  Assuming that noise in gradient measurements is isotropic and uniform within the block $B$, with variance $\sigma^2_N$ in each gradient component, its effect is the addition to the diagonal in the gradient covariance
\begin{equation}
\left(
\begin{array}{cc}
\widetilde{\sigma}_{xx} & \widetilde{\sigma}_{xy} \\
\widetilde{\sigma}_{xy} & \widetilde{\sigma}_{yy}
\end{array}
\right)
\rightarrow
\left(
\begin{array}{ll}
\widetilde{\sigma}_{xx} + \sigma^2_N & \widetilde{\sigma}_{xy} \\
\widetilde{\sigma}_{xy} & \widetilde{\sigma}_{yy} + \sigma^2_N
\end{array}
\right)
\end{equation}
which affects only the trace $\mathcal{I}_1$, but not $\widetilde{\sigma}_{xx}- \widetilde{\sigma}_{yy}$ nor $\widetilde{\sigma}_{xy}$, or, therefore, $\mathcal{J}_2$. Thus, we reach an important conclusion that our angle estimator is insensitive to, at least, such an idealized noise. Meanwhile, degree of polarization, $p_B$, is affected by noise via $\widetilde{\mathcal{I}}_1$, which decreases $p_B$ with noise addition. Thus, our estimator overerestimates true $p_B$ if the data is assumed noiseless. This, however, 
can be  corrected for, if the noise level $\sigma^2_N$ is known, but for this correction one needs to determine the trace of the gradients covariance $\widetilde{\mathcal{I}}_1$ by a separate analysis. Actually, if the trace have been determined, one can has full $\tilde{\sigma}_{ij}$. Moreover, if the full noise covariance is estimated in the data, it can be corrected for a general non-isotropic noise contribution.

In the second approach, we use a phenomenological fitting function inspired by
\cite{YL2017-HI-ApJ...837L..24Y} that 
consists of the Gaussian and the flat component
in the range of angles $[-\pi/2,\pi/2]$
\begin{equation}
\label{eq:Gaussian-fitting}
P(\theta_p) = A \exp[-(\theta_p - \tilde{\theta}_B)^2/(2 \sigma^2)] + C
\end{equation}
The model contains three independent parameters, i.e. $\bar \theta, \sigma$ and $A$,
with the fourth one, $C$, fixed by the normalization. 
Evaluating the averages $\overline{\cos 2 \theta}$ and $\overline{\sin 2 \theta}$, shows that $\theta_B = \tilde{\theta}_B$. The expression for $p_B$ is also readily obtainable, but is rather cumbersome to present it here. 

In Figure~\ref{fig:histograms}, we demonstrate the performance of both techniques. The pixels with relatively high degree of polarization (upper panels) exhibit well pronounced preferred direction with low uniform baseline component. The pixels of low polarization degree
(lower panels) show almost uniform angle distribution within, with some fluctuations over it. We can conclude that degree of polarization also serves a role
of the measure of how accurately the direction in the coarse grained block is determined.

The first method described fits the measured distributions extremely well in all cases,
for high and low polarization degree. This, in particular, supports the idea that
the gradients are nearly Gaussian distributed. The second, phenomenological choice of the
fitting function performs very well and close to the first one for relatively high 
degree of polarization pixels. However, for low polarization pixels, where the direction is determined with high degree of uncertainty to start with, there are differences.
Note that the constant term in Equation~\ref{eq:Gaussian-fitting}, while affecting 
the fit, does not contribute to the averages $\overline{\cos 2 \theta}$ and $\overline{\sin 2 \theta}$, and thus the resulting
direction or polarization degree. These two quantities only come from the Gaussian
term. With only one such term available for fit, the phenomenological formula effectively picks the most represented $\theta_p$, eliminating the contribution of any other fluctuations in angle histogram, akin of trying to find the direction which has the largest signal to noise. The first approach, on the other hand, fits faithfully variations in angle distribution and gives different result for near uniform distributions. As our comparison with observational data further in the text shows, phenomenological fits seems to perform marginally better in low polarization areas in determining the local direction $\theta_B$.  There is almost no difference between two methods for the polarization degree $p_B$.

\subsection{Relation to earlier studies}

The idea of Velocity Channel Gradients (VChGs) as a means of tracing magnetic
field was proposed in \cite{LY2018-Channel-ApJ...853...96L}.  This technique has been successfully applied to studies magnetic field in diffuse atomic hydrogen as well as in molecular clouds (see \citealt{Diego2019ApJ...880..148G}, \citealt{Hu2019NatAs.tmp..334H}). As an independent development, \citeauthor{Clark2015PhRvL.115x1302C} (\citeyear{Clark2015PhRvL.115x1302C}, \citeyear{Clark2018ApJ...857L..10C})
addressed a different problem of predicting polarized radiation from aligned dust using the technique of tracing HI intensity filaments within velocity channels. Reconstructed polarization maps can then be related to the magnetic field orientation.

Our modified VChG method shares some common features with these previous techniques,
meanwhile making some improvements. In this section, we point out what we take from the  previous work and what is different in our approach.
 
In the most general, and somewhat schematic sense, all methods of
reconstruction of the magnetic field direction $\theta(\mathbf{X})$
from the velocity channel intensity $I(\mathbf{X},v)$ have the structure:
\begin{equation}
\theta(\mathbf{X}) \sim \int dv \int d \mathbf{X^\prime}
\widehat{\mathcal{L}}_\theta \left( \mathbf{X},\mathbf{X^\prime},v\right)
\int_{\delta v} \!\! dv^\prime W\left(v,v^\prime\right)
I\left(\mathbf{X^\prime},v^\prime\right) ~,
\end{equation}
which consists of, right to left, a) assembly of the intensities in
the synthetic velocity channel of the width $\delta V$ and with weight
$W$\footnote{the width may be the whole line, the weight may include velocity
itself, e.g., $W \sim v \delta(v-v^\prime)$, which gives velocity centroids,
etc}, b) action of an operator $\hat L$ on resulting intensity map, where
$\hat L$ may be linear or non-linear, local or non-local, but is always
anisotropic, carrying information on how the intensities reflect the direction
of the magnetic field, and c) the final assembly over all the channels to
obtain the sky map of angles (and/or polarization).
 
 Among the steps above, the step b) is the central one, since that is where directional information is extracted.  This paper develops the original idea of gradient technique of
 \cite{GL2017ApJ...835...41G}, where $\widehat{\mathcal{L}}$ operator is
factorized into evaluating the angle of the local gradient of the intensity map 
and subsequent averaging over some coarse grained resolution.  Thus, the
directional information is local, while further smoothing is direction-agnostic.
Alternative notable technique of \cite{Clark2015PhRvL.115x1302C} is based on 
Rolling Hough Transform (RHT), which in the disk of a given radius around every
point on the map evaluates the radial integrated intensity as the function of
direction, after the map was treated by high-pass filter, and some thresholding
is applied. This procedure for determining direction is fundamentally non-local.
Both approaches are non-linear in intensity, due to angle evaluation from the
gradients, or intensity thresholding in RHT approach.
 
In the VChG of \cite{LY2018-Channel-ApJ...853...96L}, the raw high resolution
channel intensity maps, $I (\textbf{X},v)$, after first smoothed with a Gaussian
filter, are coadded over channels within the range $\delta v_R$ around the
average line center velocity $v_0$
\begin{equation}
\label{eq:Ch of raw VChG}
I_p(\mathbf{X})=\sum_{v_0-\delta v_R/2}^{v_0+\delta v_R/2} I_p (\mathbf{X},v)
\end{equation}
The $\delta v_R$ is chosen to be the \textit{rms} velocity at spatial scale R,
which makes the channel map under ``thin channel regim'' as defined in
\cite{LP2000ApJ...537..720L}. Then,  gradients are calculated similarly to
Equation~\ref{eq:SimpleGradient} and Gaussian fitting to their distributions within coarse grained blocks is applied to determine the coarse grained map of angles $\theta_{VChG}(I,J)$. 

We introduce the following changes relative to the original formulation
\begin{enumerate}
    \item 
    Instead of using integrated channel map, our new method calculate gradients using every channel map at raw spectroscopic resolution. According to \cite{LP2000ApJ...537..720L}, thinner the channel is, more is it affected by the velocity fluctuations, which contribution, in turn, is more sensitive to anisotropic nature of MHD. As was numerically checked by \cite{LY2018-Channel-ApJ...853...96L}, thinner channels result in higher tracing alignment measure of magnetic fields, then propagating through pseudo Stokes parameters. With much thinner channel maps used for calculation, it is natural to expect a better tracing of magnetic fields.
    \item Two methods are using different way to estimate gradients direction inside sub-blocks. The original VChG method fit the angle distribution by a Gaussian function of Equation~\ref{eq:Gaussian-fitting} and choose the most probable angle to represent the coarse pixel. Whereas our new method focuses on evaluating mean values of $cos(2\theta_p)$ and $sin(2\theta_p)$ to define pseudo Stokes parameters for the coarse pixel. For high precision, this requires the phenomenological Gaussian fit (which is not periodic) to be performed iteratively adjusting the periodic interval of angles to obtain the mode of the distribution at the center. Our theory-inspired method is based on distribution of Equation~\ref{eq:Gaussian-grads} ,which automatically respects the periodicity condition. Our approach is more robust than that of the mode of the distribution, which, in particular, allowed us use narrow velocity channels.  But more importantly, using information on the width of the angle distribution as well, we are able to depict a complete picture of the extent of angular variations inside the coarse pixel, which allows us to robustly estimate its degree of polarization. 
\end{enumerate}

The idea of extracting anisotropy at each velocity channel and propagate it by
pseudo Stokes parameters was introduced to the field by \cite{Clark2015PhRvL.115x1302C} and improved by \cite{Clark2018ApJ...857L..10C}. We similarly found this approach to be natural and useful, especially if the target is to compare with dust polarization maps, and we utilize it as well. Once $\tilde Q$ and $\tilde U$
are defined by angle averaging in RHT weighted pixel of \citep{Clark2018ApJ...857L..10C} or in our coarse-grained pixel, the subsequent treatment of polarization is the same. Note, however, that if the goal is to obtain a direct reconstruction of the magnetic field orientation, the pseudo-Stokes approach is only one of possible ways to weight the
orientation information in individual channels, which remains to be studied whether it is the most optimal one.

The main difference between our method and \cite{Clark2018ApJ...857L..10C} method
is that we evaluate the anisotropy in individual channel by simply calculating
linear gradients and then averaging in a square block; she evaluates the anisotropy by applying RHT to extract filament structures. Both methods can produce pseudo Stokes parameters, whereas our linear gradient kernel is much lighter than the RHT kernel, where a lot of computational resources can be saved.

Moreover, the simplicity of our operator allows for a more straightforward
theoretical analysis of its properties, as started in this paper. In particular,
a notable difference comes from the local nature of our angle estimator versus RHT.  With the knowledge of the gradient angular distribution within super-blocks, we are able to reveal more physical information, for example, Alfvenic Mach number (\citealt{Lazarian2018-MA-ApJ...865...46L}; \citealt{Hu2019NatAs.tmp..334H}).
Locality also gives us a straightforward way to include in the modelling the variations of the degree of polarization along the individual lines of sight that combine in the coarse grained pixel. These variations might
reflect three dimensional fluctuations in the magnetic field along the line of sight, which is important for a more accurate modelling.

\section{Numerical test in different magnetization}
\label{sec:numeric}

\begin{figure*}
\centering
\includegraphics[width=1.0\textwidth]{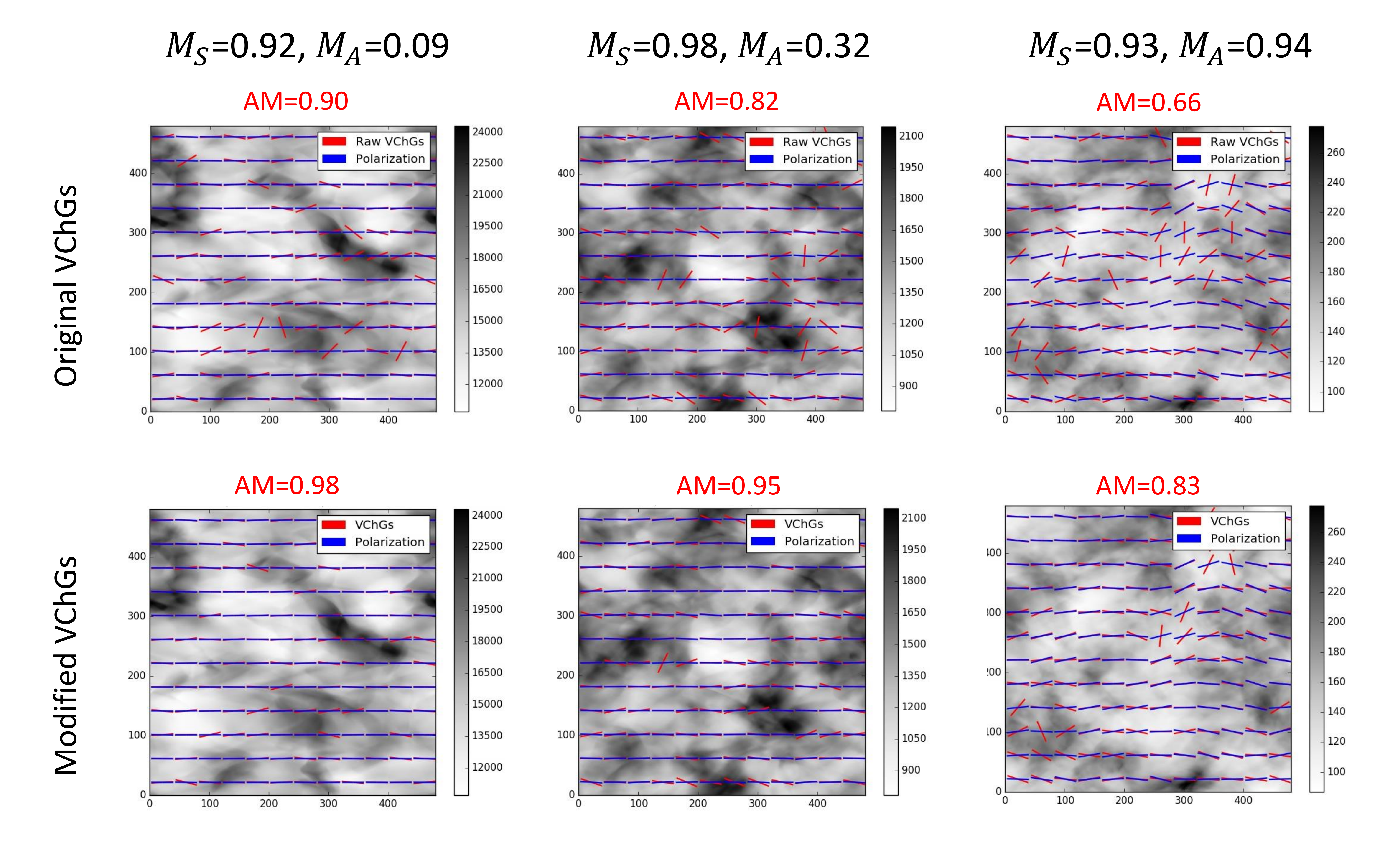}
\caption{Comparison of tracing performance of the original VChG and the modified VChG in subsonic regime, for several Alfvenic Mach numbers. The blue segments show the simulated orientation of dust polarization, and the red segments show the orientation of dust polarization predicted by different VChGs. The parameters of simulations are from \autoref{tab:beta}.}
\label{fig:subsonic}
\end{figure*}

\begin{figure*}
\includegraphics[width=1.0\textwidth]{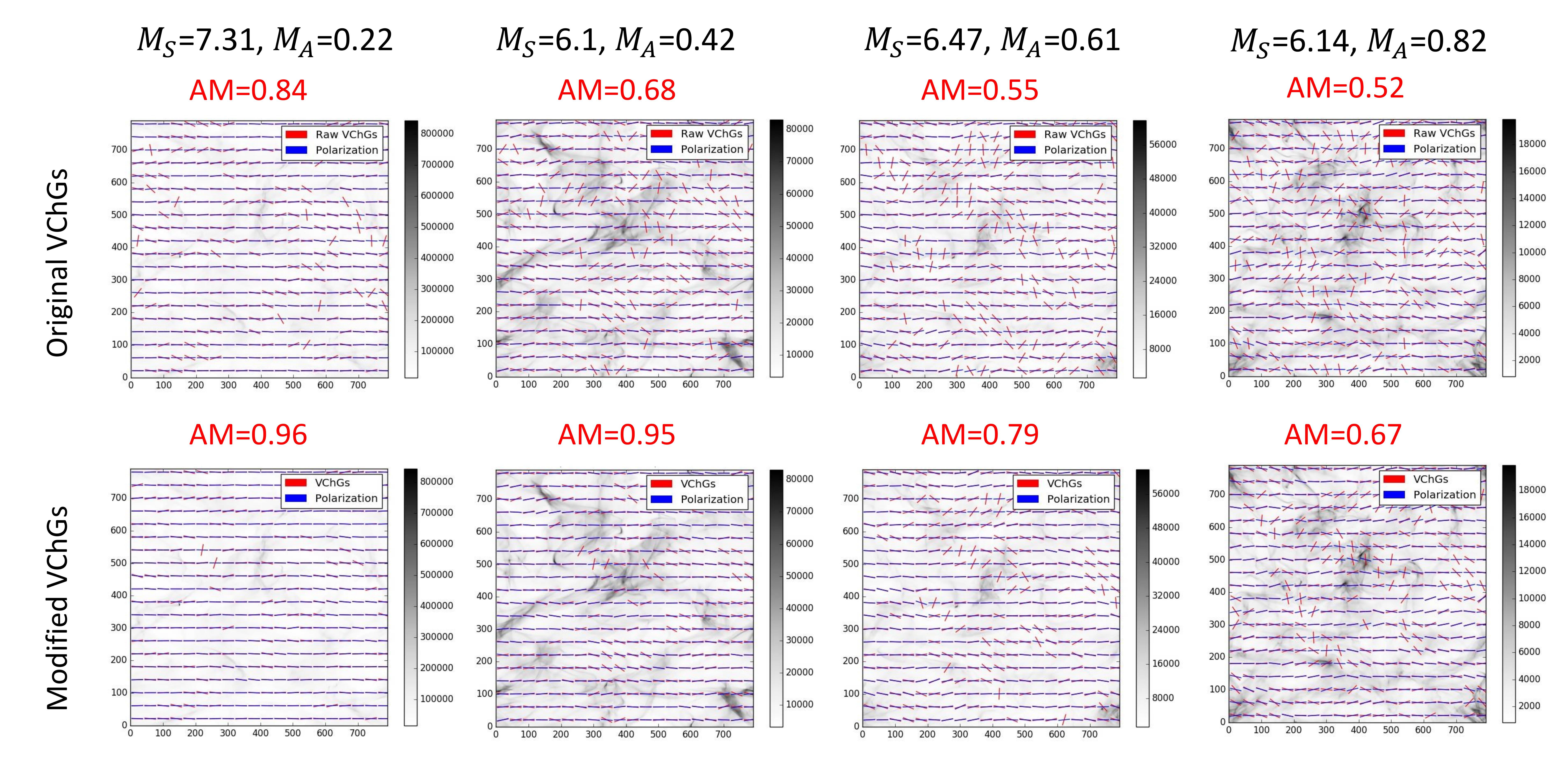}
\caption{Comparison of tracing performance of the original VChG and the modified VChG in supersonic regime, for several Alfvenic Mach numbers. The blue vectors show the simulated orientation of dust polarization, and the red vectors show the orientation of dust polarization predicted by different VChGs. The parameters of simulations are from \autoref{tab:huge}.}
\label{fig:supersonic}
\end{figure*}

As a new method to trace magnetic fields, we are interested in two things: the comparison with the original version of the VChG in terms of polarization orientation tracing; VChG' performance in media with different magnetization, namely, different Alfvenic Mach number. To do this, we test our method in simulation sets.

In single fluid MHD simulations, we simulate three-dimensional velocity and density of the medium, as well as magnetic fields vectors. To predict dust distribution, we assume that it follows the simulated fluid.
Using the dust model given by \cite{Wardle1990ApJ...362..120W}, we can connect dust polarization with magnetic fields by computing Stokes parameters as
\begin{equation}
\label{eq:Dust_Stokes_Mag}
\begin{aligned}
Q_{dust}(x,y)& = p' \int \rho(x,y,z)\ \frac{B_x^2-B_y^2}{B_x^2+B_y^2+B_z^2}\ dz ~,\\
U_{dust}(x,y)& = p' \int \rho(x,y,z)\ \frac{2B_xB_y}{B_x^2+B_y^2+B_z^2}\ dz ~,
\end{aligned}
\end{equation}
where $\rho$ is the density; $B_x, B_y, B_z$ are magnetic field components, and
it is assumed that individual dust grains have a fixed degree of polarization
$p'$. Such fixed value is not important if we consider only the orientation of polarization 
$\theta_{dust}=\frac{1}{2}\mathrm{atan}\frac{U_{dust}}{Q_{dust}}$,
as we do in this section.  If we look in detail how Stokes parameters
accumulate along the line of sight 
\begin{equation}
\label{eq:Dust_Stokes_Mag_in angles}
\begin{aligned}
Q_{dust}(x,y)& = p' \int \rho(x,y,z)\ \cos 2\theta(x,y,z) \sin^2\psi(x,y,z) \ dz ~,\\
U_{dust}(x,y)& = p' \int \rho(x,y,z)\ \sin 2\theta(x,y,z) \sin^2\psi(x,y,z) \ dz ~,
\end{aligned}
\end{equation}
where $\theta$ is the local sky orientation angle
and $\psi$ is the angle with
the line-of-sight of the magnetic field. We should note that $p'$ represents the degree of polarization of emitted individual dust grains, which we assume to be a constant. Whereas, the observed degree of polarization, $p_{dust}(x,y)=\sqrt{Q_{dust}^2+U_{dust}^2}$ will, in general,
differ for different lines of sight. 

The simulation sets in this section are the same as in \cite{LY2018-Channel-ApJ...853...96L} and \cite{Lazarian2018-MA-ApJ...865...46L}, which adapted a series of compressible, turbulent, isothermal single fluid magnetohydrodynamic (MHD) simulations from ZEUS-MP/3D, a variant of the well-known code ZEUS-MP (\citealt{Norman2000PhRvB..6114751N},  \citealt{Hayes2006ApJS..165..188H}). The subsonic cubes ($M_s \sim$ 0.9, $M_A$ ranges from 0.09 to 0.94, see \autoref{tab:beta}) with relatively low resolution, $480^3$, allow us to take quick tests of our recipe. Meanwhile,  the supersonic cubes ($M_s\sim$ 7, $M_A$ ranges from 0.22 to 0.82, see \autoref{tab:huge}) with resolution of $792^3$, allow us to test our recipe more accurately. Note that the Alfvenic Mach number is defined as $M_A=V_L/V_A$; the sonic Mach number is defined as $M_S=V_L/V_s$, where $V_L$ means the turbulence injection velocity, $V_A$ means the Alfvenic velocity, and $V_s$ means the sonic velocity.

\begin{table}
\centering
 \begin{tabular}{c c c c}
 \hline
 Model & $M_s$ & $M_A$ & Resolution\\ \hline 
   Ms0.8Ma0.08 & 0.92 & 0.09 & $480^3$ \\
   Ms0.8Ma0.264 & 0.98 & 0.32  & $480^3$ \\
   Ms0.8Ma0.8 & 0.93 & 0.94 & $480^3$ \\  \hline
\end{tabular}
 \caption {Parameters of subsonic MHD simulation sets used. The Mach numbers in column ``Model'' are the initial values for a simulation. $M_s$ and $M_A$ are the instantaneous values at final snapshots. Resolution of the simulated cubes is $480^3$.}
 \label{tab:beta}
\end{table}

\begin{table}
\centering
\begin{tabular}{c c c c c}
\hline
Model & $M_S$ & $M_A$  & Resolution \\ \hline 
Ma0.2 & 7.31 & 0.22 &  $792^3$ \\  
Ma0.4 & 6.1 & 0.42 &  $792^3$ \\
Ma0.6 & 6.47 & 0.61 &  $792^3$ \\  
Ma0.8 & 6.14 & 0.82 &  $792^3$ \\  
 \hline 
\end{tabular}
\caption{Parameters of supersonic MHD simulation sets used. The Mach numbers in column ``Model'' are the initial values for the simulations. $M_s$ and $M_A$ are the instantaneous values at final snapshots. Resolution of the simulated cubes is $792^3$.}
\label{tab:huge}
\end{table}

As described in section \ref{sec:Method}, we put forward the modified version of VChG by imitating the addition process of Stokes parameters, and keeping velocity channel at their highest resolution.  Hence, we expect a higher tracing precision of polarization when compared to the original VChG. To obtain a simple quantitative characterization of the angle difference $\Delta \phi$ between two maps covering the same region, we define the alignment measure (AM) as follow:\footnote{The alignment measure of this type was first introduced in Gonsalvez-Casanova \& Lazarian (2017) and used in the subsequent papers. The alignment can be measured with this measure in respect to magnetic field or with polarization that acts as the proxy of the projected magnetic field. As we discussed earlier, due to the different properties of magnetic fields and polarization, as far as adding along the line of sight is concerned, the direction of the polarization obtained obtained by the averaging along the line of sight may be different from the averaged along the line of sight direction of magnetic field. However, these differences are not important within our present discussion. }
\begin{equation}
\label{eq:AM}
AM=2\ \langle \cos^2(\phi) - 1/2\rangle \equiv \langle \cos 2 \phi \rangle
\end{equation}
where the $\phi$ is the angle difference between two vectors and averaging is done over the whole region.
AM is a value ranging from -1 to 1, which provides us with an quantification of the overall alignment: AM=1 means perfect alignment; AM=0 means no relationship; AM=-1 means two fields vector are perfectly perpendicular. Note that the measure does not distinguish between forward and backward pointings of the vectors, depending only on the axis of direction.

The comparison between synthetic polarization orientation and simulated polarization orientation is shown in Figure~\ref{fig:subsonic} (subsonic, $M_s \sim$ 0.9) and Figure~\ref{fig:supersonic} (supersonic, $M_s\sim$ 7). From the two figures, we can see that
\begin{enumerate}
    \item AM of VChGs tend to decline with increasing $M_A$.
    \item Tracing precision of the the pseudo-Stokes parameters based VChG (modified VChG) is improved versus the original version of VChG, $AM > 0.8$ even for $M_A \sim 0.9$.        
\end{enumerate} 
Note that since the original VChG did not predict degree of polarization, we only compare the orientation of polarization here. For orientation, the new version of VChG improves the tracing precision and works well in all sub-Alfvenic $M_A < 1$ regimes that were tested. These conclusions are robust in both subsonic and supersonic cases.

\section{Applying to observation}
\label{sec:observation}

\begin{figure*}
    \centering
    \includegraphics[width=0.9\textwidth]{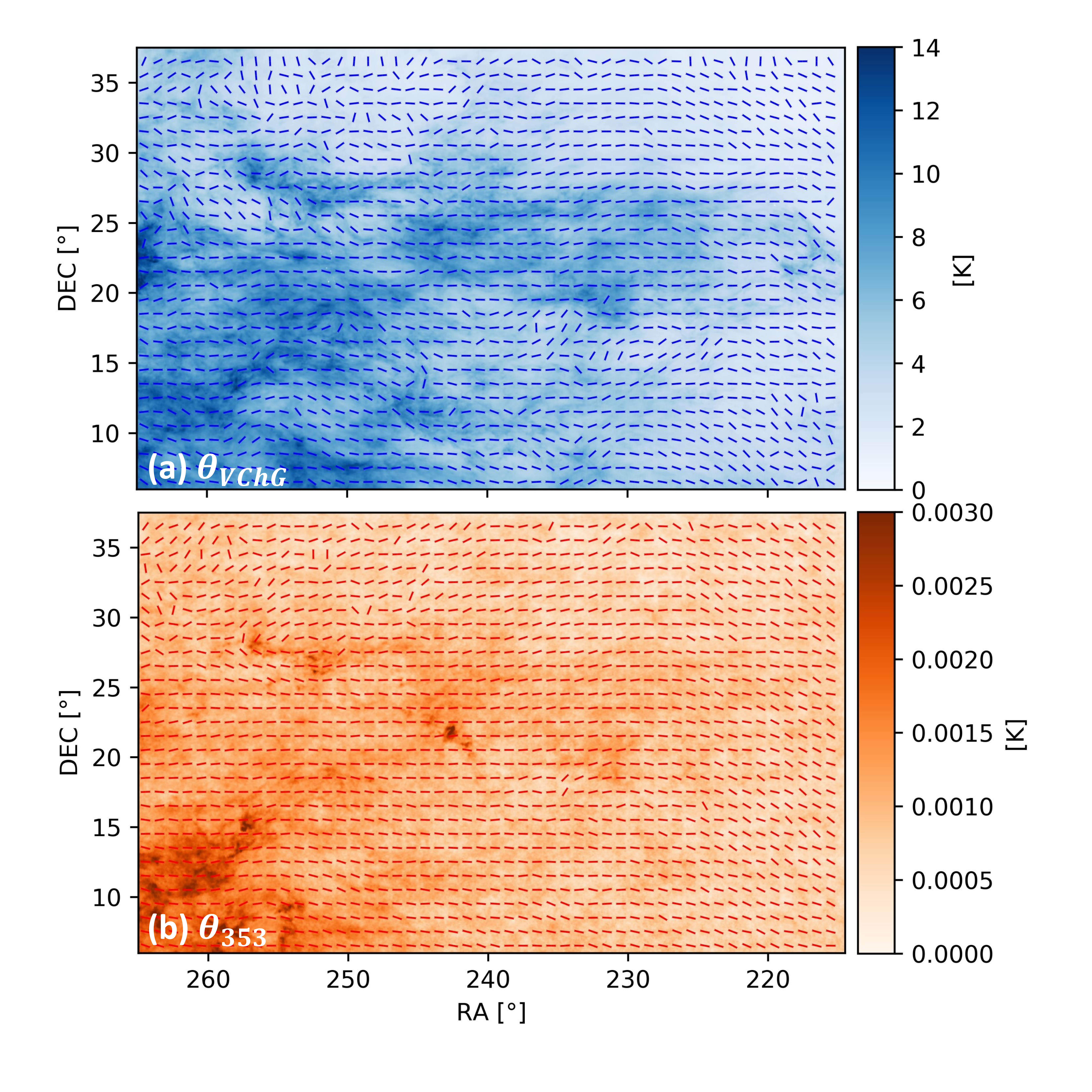}
    \caption{Polarization direction maps of the Planck 353 GHz (panel (a)) and the VChGs (panel (b)), where the background maps are the projected plane-of-sky intensity of GALFA-HI (blue) and Planck (orange). The VChG is calculated by GAFLFA-HI data with 0.184 km/s velocity channel width, ranging from -13.52 km/s to 13.52 km/s. Both the direction maps are pixelized in resolution of 1\degree; the background intensity maps are pixelized in their raw resolution of GALFA-HI and Planck.}
    \label{fig:pol}
\end{figure*}

\begin{figure*}
    \centering
    \includegraphics[width=0.9\textwidth]{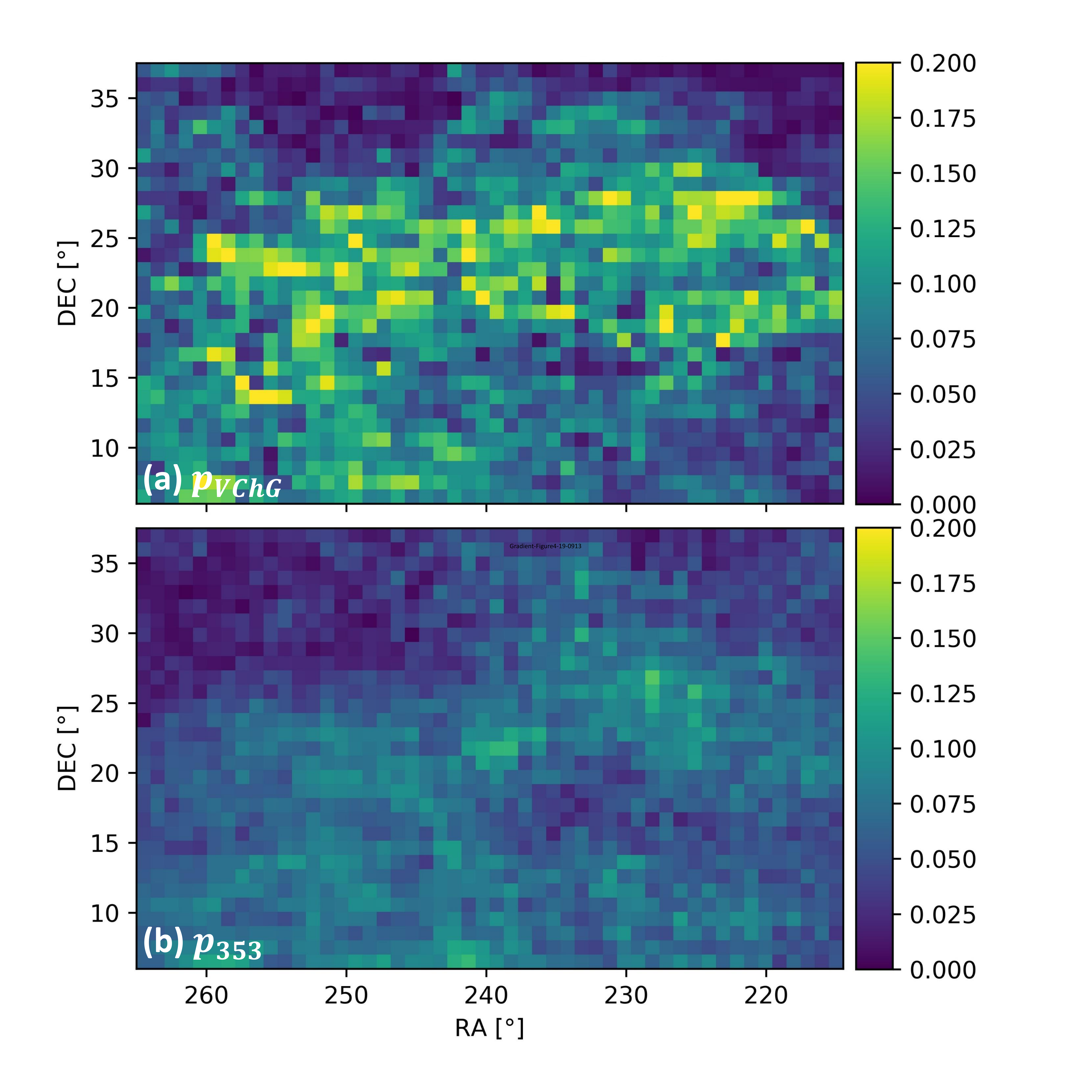}
    \caption{Maps of the degree of polarization of the VChG (panel (a)) and the Planck 353 GHz (panel (b)), where the spatially mean values $\langle p_{VChG} \rangle=0.082$ and $\langle p_{353} \rangle=0.059$. Both maps are pixelized in resolution of 1\degree. }
    \label{fig:pp}
\end{figure*}

\begin{figure*}
    \centering
    \includegraphics[width=0.9\textwidth]{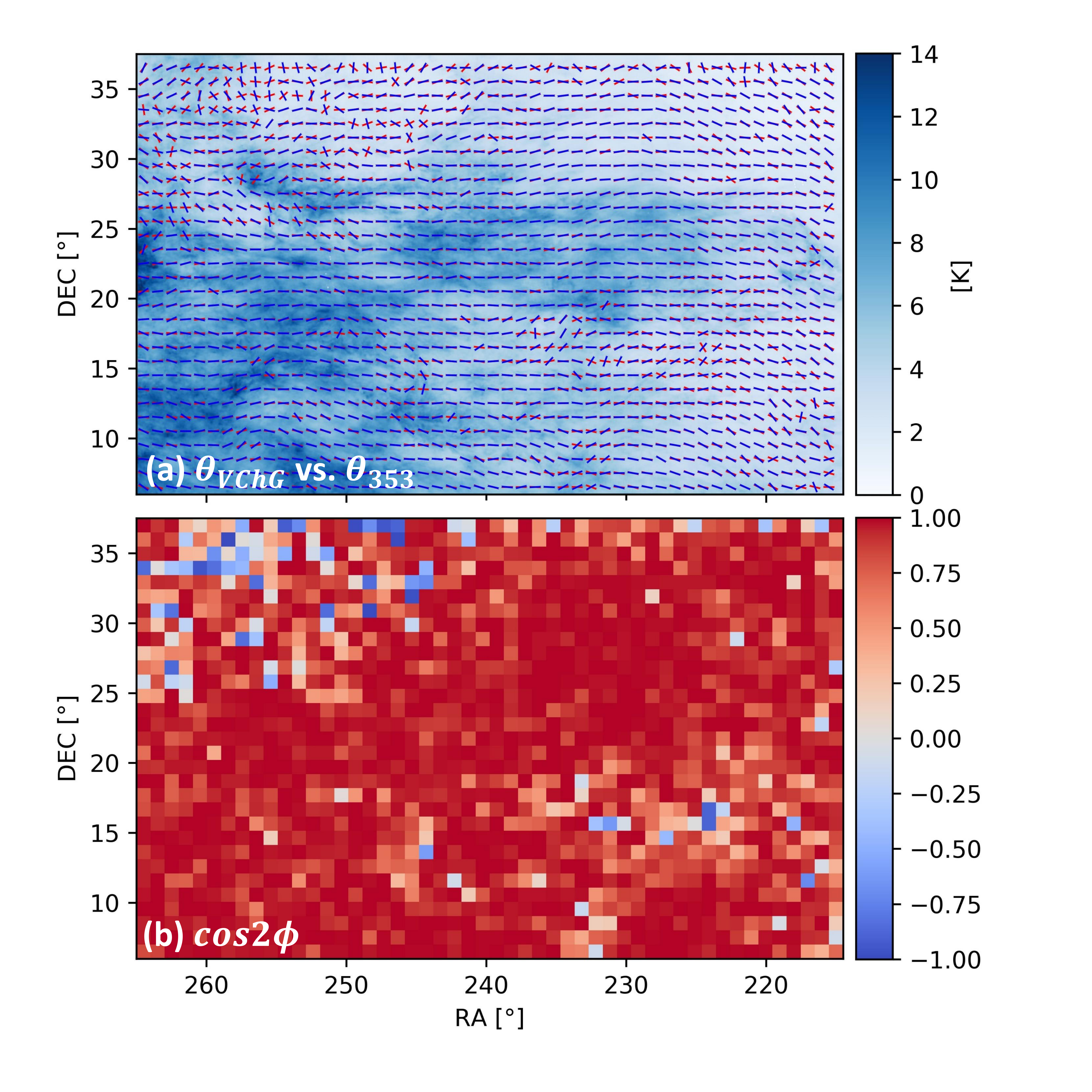}
    \caption{The panel (a) shows the pixel-by-pixel comparison between $\theta_{VChG}$ (blue) and $\theta_{353}$ (red), where the background intensity map is the projected HI intensity. The panel (b) shows the spatial distribution of $cos2\phi$, whose spatially mean value $AM=0.77$. Note that $\phi$ is the angle difference between $\theta_{353}$ and $\theta_{VChG}$. Both maps are pixelized in resolution of 1\degree.}
    \label{fig:cos2theta}
\end{figure*}

\begin{figure*}
    \centering
    \includegraphics[width=0.9\textwidth]{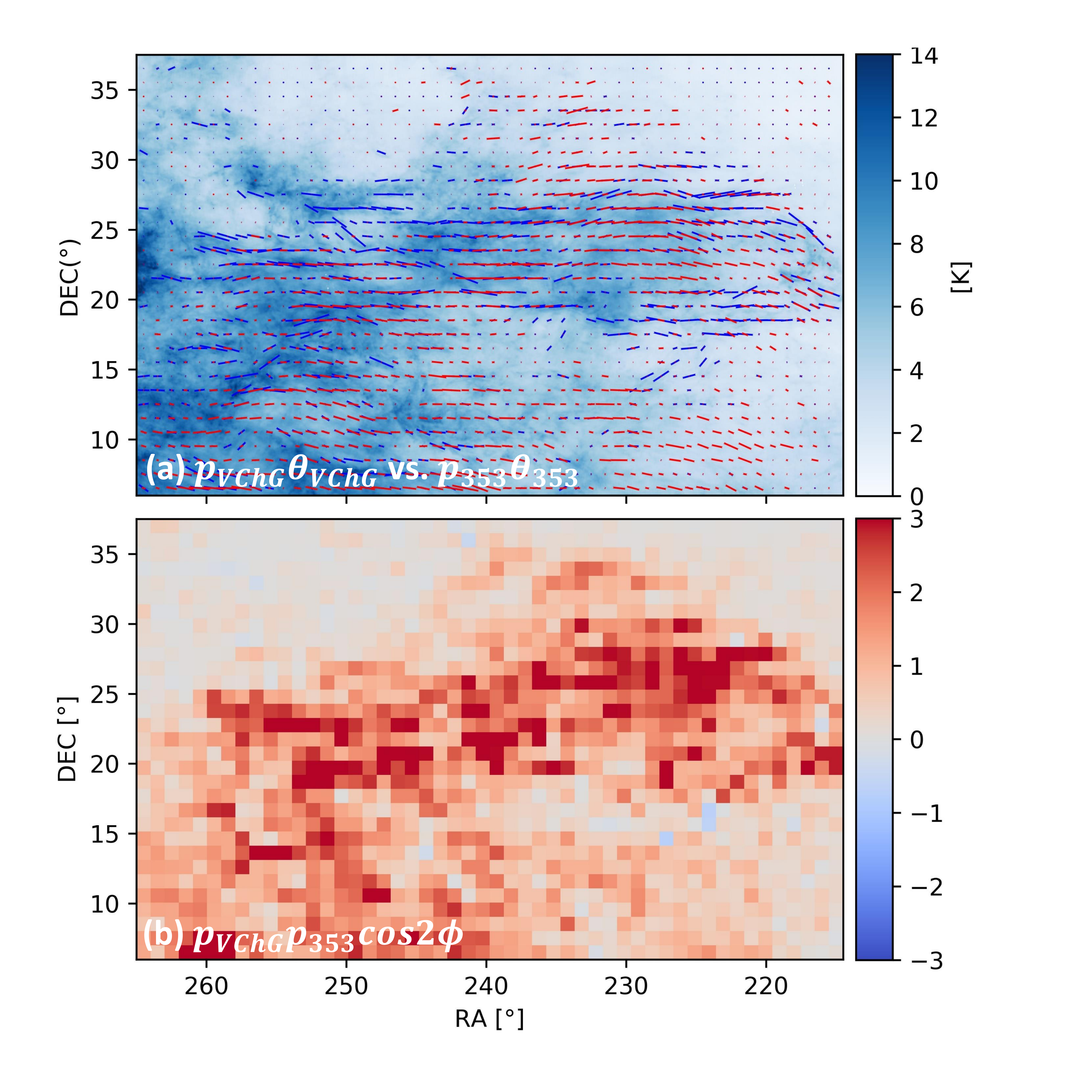}
    \caption{The panel (a) shows the pixel-by-pixel comparison between $p_{VChG}\theta_{VChG}/\langle p_{VChG}\rangle$ (blue) and $p_{353}\theta_{353}/\langle p_{353}\rangle$ (red), where the background intensity is the projected HI intensity map. The panel (b) shows the map of $p_{VChG}p_{353}cos2\phi/\langle p_{VChG}p_{353}\rangle$, whose spatially mean value $ppAM=0.89$. Note that $\phi$ is the angle difference between $\theta_{353}$ and $\theta_{VChG}$. Both maps are pixelized in resolution of 1\degree.}
    \label{fig:ppcos2theta}
\end{figure*}

\begin{figure}
    \includegraphics[width=0.5\textwidth]{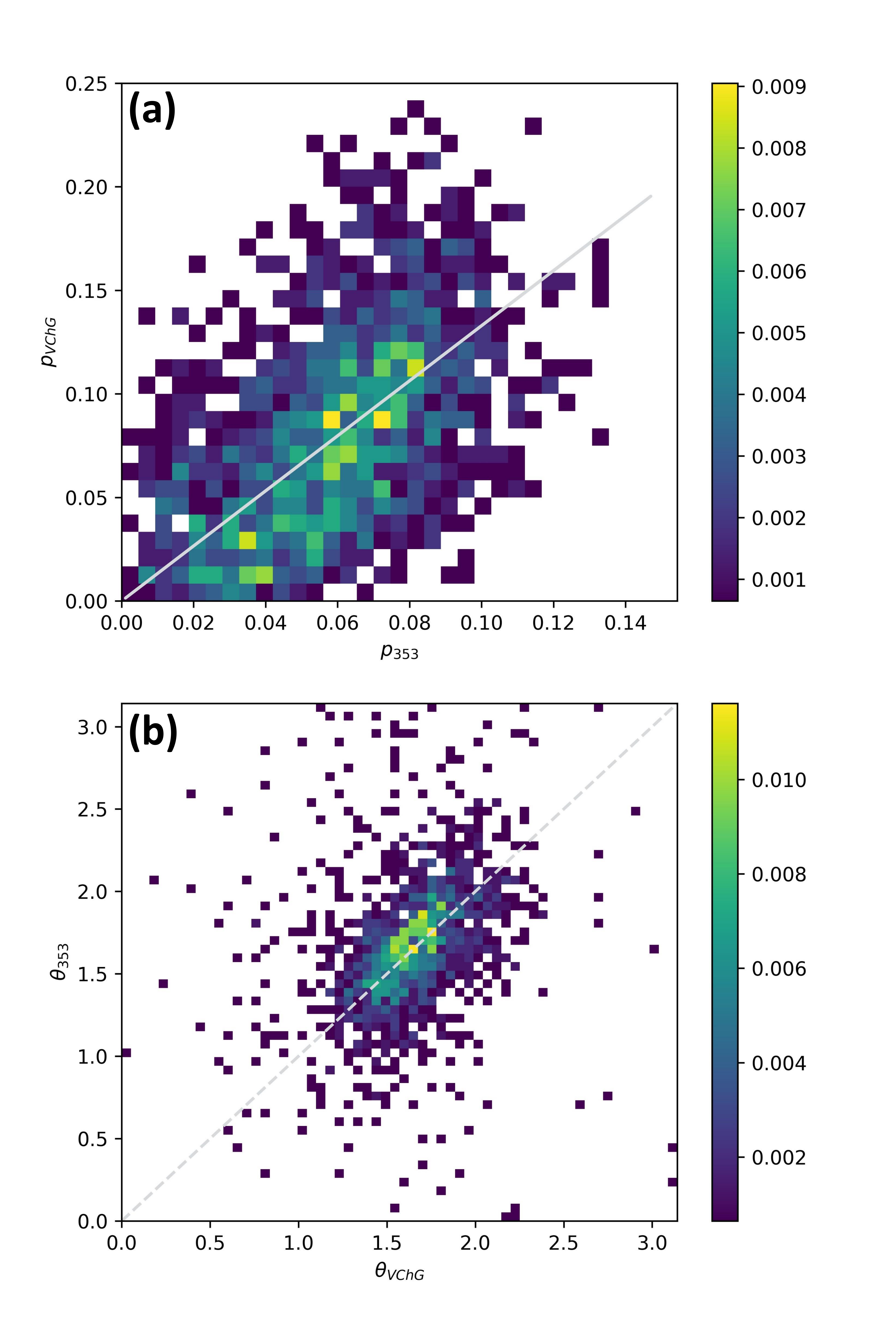}
    \caption{The panel (a) shows the two-dimensional histogram of the degree of polarization between Planck 353GHz ($p_{353}$) and VChG ($p_{VChG}$), where solid grey line is the least square fitting of the their scatters, using our one-parameter linear model. The slope of fitting line is 1.33. The panel (b) shows the two-dimensional histogram of the polarization orientation of VChG ($\theta_{VChG}$) and Planck 353GHz ($\theta_{353}$). The dashed line is a reference line $\theta_{VChG}=\theta_{353}$ ,along which the alignment is perfect. Both histograms are normalized by the total count.}
    \label{fig:hist2d}
\end{figure}

With the positive conclusions obtained in section~\ref{sec:numeric}, next we apply VChG to real observations. 
We produce the synthetic map of dust polarization from HI data in velocity channels, for which we use GALfA-HI DR2 \citep{Peek2018ApJS..234....2P}.
To test the accuracy of our synthetic map, we will compare it with the Planck's observed polarization at 353 GHz \citep{Planck_Collaboration2015A&A...576A.104P}. 

\subsection{Data}
We use recently released HI data from GALfA-HI DR2 (The Galactic Arecibo L-band feed Array HI, see \citealt{Peek2018ApJS..234....2P}), a survey of the 21 cm HI line over the Arecibo sky (decl. $−1\degree17'$ to $+37\degree57'$across all R.A.) at $1' \times 1'$ spatial resolution and 0.184 km/s spectral resolution (``Narrow'' set), with 150 mK median rms noise per 1 km/s channel. For our analysis, we choose the sky region of GALfA-HI data whose right ascension (R.A.) ranges from $215.0\degree$ to $265.0\degree$ and declination (DEC.) ranges from $6.0\degree$ to $37.5\degree$, to avoid
the regions near the Galactic plane and the North Galactic pole. Our region
is similar to the field studied in \cite{Clark2018ApJ...857L..10C} and twice
larger in declination range to the field used for magnetic field comparison
in \cite{Clark2015PhRvL.115x1302C}. 
We also use the velocity channels spanning the range from $-13.52$ km/s to $13.52$ km/s.   
Following the steps described in \autoref{sec:Method}, we apply VChG to the PPV cubes of HI, during which thin channels, 0.184 km/s, are used and a $1\degree\times1\degree$ block average is applied. The VChG angles and ``degree of polarization'' from HI data are calculated from Equations~\ref{eq:VChG_theta},
\ref{eq:VChG_p}.

As a comparison with the synthetic dust polarization predicted by VChG, we plot the dust polarization observed by Planck mission. Since 353 GHz emission is dominated by the thermal dust, we use single frequency 353 GHz polarization map by PLANCK satellite to plot dust polarization (see \citealt{Planck_Collaboration2015A&A...576A.104P}). 
Using CMB-cleaned 353 GHz map or the component separated dust map, also provided by the Planck team, lead to no change in the conclusions.

Planck maps are provided in HEALPix\footnote{\url{http://healpix.sourceforge.net}} pixelization
at $Nside=2048$ which corresponds to approximately $1.7'$
pixel linear size. The following processing is applied.
Firstly, we smooth the observed Stokes parameters, $I_
{353}$, $Q_{353}$ and $U_{353}$ by a Gaussian function with a $5'$ FWHM. Secondly, to compare with GALFA-HI data, we transform the $Planck$ Stokes parameters to Equatorial coordinates, and select the GALFA field. Thirdly, we
rebin Planck data using HEALPix 
provided interpolating functions onto a Cartezian grid with $1\degree \times 1\degree$ pixels, by averaging $I_
{353}$, $Q_{353}$ and $U_{353}$ within the coarse pixels. The polarization angle $\theta_{353}$
and the degree of polarization $p_{353}$ are then obtained by the usual
relations to $I_{353}$, $Q_{353}, U_{353}$.

\subsection{Results}
\label{subsec:Results}

The maps of polarization directions from VChG and Planck 353GHz are shown in panel (a) and panel (b) in Figure~\ref{fig:pol} respectively. The orientation of the line segments (here of equal length) shows the direction of the polarization; the background intensity maps show the brightness temperature of HI 21cm emission and intensity of 353 GHz emission. The direction map allows us to visualize the underlying direction of the magnetic field. In the case of VChG, the surface brightness map is integrated along all velocity 
channels, so the gradients in individual channels that the VChG method is based upon are not explicitly the gradients of the intensity shown, though there is some correspondence.

In Figure~\ref{fig:pp}, we show respectively the reconstruction of the degree of polarization by VChG in panel(a) and the degree of polarization map for Planck 353~GHz in panel(b). We see significant visual similarities in the structure of $p$ maps. At the same time, VChG results give level of polarization that is higher than the Planck data. This is under expectation, since in VChG we did not account for only partial polarization level of individual dust grain emitters to begin with, as well as depolarization due to fluctuations of the magnetic field along the line of sight when the emission was collected in a channel from different physical depths. We note that simple constant scaling, e.g. with the ratio of
the spatial means of two maps $\left\langle p_{VChG}\right\rangle/\left\langle p_{353}\right\rangle \approx 1.39$, is insufficient to explain all the differences. This points to the necessity to model variation of polarization degrees $o_i$  between individual lines-of-site to achieve better correspondence.  Gradient technique advanced in this paper has advantage of being able to easily incorporate such models by assigning Stockes parameters to fine-grain pixels, with local direction given by the gradient as before, but using non-uniform $p_p(i,j)$ before combining the data into coarse-grained $\widetilde{Q}_B, \widetilde{U}_B$.

Let us now turn to the quantitative comparison between two datasets.  In Figure~\ref{fig:cos2theta} we compare only the directional information. The top panel overlays the two maps from Figure~\ref{fig:pol} to see the matching and the differences in
pixel to pixel direction of polarization, while the bottom panel is the explicit map of angle difference measure 
$\cos 2 (\theta_{VchG} - \theta_{353})$. We see that over most pixels of the map this measure is close to unity, indicating a high degree
of alignment. Indeed, the average of the whole map value is the earlier introduced $AM = \left\langle \cos 2 \phi \right\rangle$,
where $\phi=\theta_{353} - \theta_{VChG}$. We obtain $AM = 0.77$ for VChG as compared to Planck 353 Ghz. 
This level of accuracy of reconstructing the direction from HI data is
on par with the best results quoted in the literature, e.g. \cite{LY2018-Channel-ApJ...853...96L}, despite the much larger area studied and simplicity of our approach.

The AM alignment measure mostly reflects the variance of angle difference between two maps.  Indeed, at least at small
angle differences
$\left\langle \cos 2 \phi \right\rangle \approx 1 + \left\langle\phi^2 \right\rangle/2 $.   It is also instructive to consider an associate measure, 
$sAM=\left\langle \sin 2 \phi\right\rangle$, which, in contrast, reflects the systematic
mean deviation of directions in one map versus another (this measure will vanish if two maps are perfectly aligned or if
the angle difference has equal probability to be of opposite signs. It is equal to maximum $+1$ if one map direction is $\pi/4$ above the other and $-1$ if it is $\pi/4$ below). The pixel map of this quantity is difficult to interpret, in particular because pixels with near perpendicular directions give equally low value to ones with parallel alignment, so we do not present it here. However, the overall average in our comparison is $sAM= 3.3\times10^{-2}$, which indicates lack of systematic angular deviation.

To include the full polarization information in our quantitative comparison, we evaluate the correlation between Planck's polarization
map and synthetic polarization map using the cross-correlation function of complex polarization $P=Q+i U$ 
at zero lag, whose pixel by pixel estimator is
\begin{equation}
\label{eq:correlation_function}
\begin{aligned}
\xi(\mathbf{X})& \equiv P_{353}(\mathbf{X})\ P^*_{VChG}(\mathbf{X}) \\
&= Q_{353}Q_{VChG}+U_{353}U_{VChG}\\
&+i \left( U_{353}Q_{VChG}-Q_{353}U_{VChG}\right)
\end{aligned}
\end{equation}
Divided by the intensities $I_{353}I_{VChG}$, we can interpret the correlation function as
\begin{equation}
\label{eq:correlation_function_parts}
\begin{aligned}
    \frac{\xi}{I_{353}I_{VChG}}&= p_{353}p_{VChG}\cos 2 \phi
    + i p_{353}p_{VChG}\sin 2\phi
\end{aligned}
\end{equation}
Therefore, the real part of $\xi$ measures the alignment between Planck's orientation and synthetic orientation weighted by the product of degrees of polarization $p_{353}p_{VChG}$. Taking the overall average and further
normalizing the weights we can define a polarization degree weighted alignment measure as
\begin{equation}
\label{eq:ppAM}
ppAM=\frac{\langle  p_{353}p_{VChG} \cos2\phi \rangle}{\langle  p_{353}p_{VChG}  \rangle}
\end{equation}
Here $ppAM$ also varies from -1 to 1, with main difference from $AM$ is that $ppAM$ focuses more on the regions with higher degree of polarization. Then, we are able to quantitatively compare the polarized direction maps between two datasets. As shown in Figure~\ref{fig:ppcos2theta}, the panel (a) overlaps the polarized vectors of VChG and Planck 353GHz, where the vectors are weighed by its degree of polarization; more polarized it is, longer the vectors will be. The Panel (b) shows the spatial distribution of $p_{VChG}p_{353}cos2\phi/\langle p_{VChG}p_{353}\rangle$, whose average value $ppAM=0.89$. A higher value of $ppAM$ than $AM$ indicates that our VChG method traces the real polarization better in more polarized regions. This is not surprising, since
high degree of polarization also reflects the low variance in gradient directions, and thus lower uncertainty in determining the polarization direction within the coarse-grained pixel block. Therefore, from another point of view, ppAM can be considered as an alignment measure where angle differences are weighted by their inverse uncertainty.
Similarly, we can evaluate the imaginary part of the correlation function, $\langle p_{VChG}p_{353} \sin2 \phi \rangle/\langle p_{VChG}p_{353}\rangle$, which equals $5.7\times 10^{-3}$, which again points to the lack of systematic misalignment.  

To inspect the correlation between the Planck dust polarization map and the VChG synthetic dust polarization map from another perspective, we plot the two-dimensional histogram between them (Figure~\ref{fig:hist2d}), in the sense of degree of polarization in panel (a) and orientation of polarization in panel (b). As for the degree of polarization, we can see a good correlation between $p_{353}$ and $p_{VChG}$. To quantify the systematic deviation of our estimation, we linearly fit the scatter using a one-parameter model, $p_{VChG}=a\ p_{353}$, where the slope $a$ means statistically the multiple that we overestimate the degree of polarization. As shown in panel (a), the result given by lease square fitting is $p_{VChG}=1.33\ p_{353}$. As for the orientation of polarization, we can also see a good correlation between $\theta_{353}$ and $\theta_{VChG}$. If the two angle statistically equal with each other, $\theta_{VChG}=\theta_{353}$, we expect a line starting from the origin with slope equalling one to describe their relation. We can see from the grey dashed line in panel (b) that such line well matches with the histogram.

\section{Discussion}
\label{sec:discussion}

Our present study is the first study exploring the applicability of the VChG to predict dust polarization. The modifications to original VChG demonstrate suggests an increasingly good correspondence with Planck 353GHz dust map. At the same time, at the current stage, the VChG also has several differences from real dust model, as well as some limitations. 

First of all, our current model does not account to full extend for three dimensional orientation of the magnetic field along the line of sight. If we look back to the real dust model (\autoref{eq:Dust_Stokes_Mag_in angles}), we can see that the accumulation of Stokes parameters depend not only on the plane-of-sky orientation of magnetic field, $\theta$, but also the line-of-sight orientation of the magnetic field. More specifically, it depends on the angle difference between line-of-sight magnetic field with the plane-of-sky, the $sin^2\psi$ term in formula. 
In our VChG method, by measuring the gradient of HI intensity on the sky, we could only predict, statistically, the projected plane-of-sky orientation of magnetic field, i.e $U/Q$ of \autoref{eq:Dust_Stokes_Mag_in angles}, but not the degree of polarization along an individual line of sight nor, even more importantly, its fluctuations between the lines-of-sight.
Such incompleteness decrease the accuracy of synthetic dust map. Improvements call for more close studies of
the three dimensional magnetic field distribution. We address this issue in the forthcoming publications.

Our model of polarization arising from aligned dust can also be improved. We assumed perfect alignment of silicate dust, which is a reasonable assumption for the radiative torque (RAT) alignment of dust grains in diffuse regions
(see \citealt{LazarianHoang2007MNRAS.378..910L}; \citealt{Andersson2015ARA&A..53..501A}). This have direct relevance 
to the issue of the normalization of parameter $p$. 
The issue of constant temperature of dust that we assumed seems to be more controversial and it requires more studies. We treat this as only first approximation to the complex problem that we address. 

At this moment, our maps can be used as a prior for removing the foreground polarization from dust. This use of HI was suggested in \cite{Clark2015PhRvL.115x1302C} and \cite{Clark2018ApJ...857L..10C}. Similar to Clark's idea, we can estimate the change of the degree of polarization. However, a big difference between our approach and that in \cite{Clark2018ApJ...857L..10C} is that we use the distribution of directions withing sub-blocks, rather than just the change of the direction of the filaments measured in different channel maps. In this sense, our approach provides more statistical information , which makes our estimates of polarization more reliable. 

We note that the nature of the intensity fluctuations in the channel maps has been debated recently. \cite{Clark2019ApJ...874..171C} maintained that these filaments are actual density filaments, while in \cite{Yuen2019arXiv190403173Y} it was claimed that velocity caustics play the dominating role for creating structures in the thin channel maps that are employed for the analysis. For our present study, this controversy is irrelevant as the gradient technique is agnostic as to interpretation of the structures in PPV space and works well with both structures created by velocities and densities (see \citealt{Hu2019arXiv190809488H}).

 The studies in \cite{Kandel2016MNRAS.461.1227K} 
 suggest that the Alfven Mach number corresponding to the high latitude dust is less than unity. This helps to our analysis, as the VChG for low $M_A$ works better and does not require additional spatial filtering of low spacial frequencies (see \citealt{LY2018-Channel-ApJ...853...96L}).

\section{Summary}
\label{sec:summary}
In this paper, we successfully improved the VChG technique to produce synthetic maps of dust polarization, both orientation and degree of polarization. To summarize, we have reached the following conclusions:
\begin{enumerate}
\item Adding up the pseudo-Stokes parameters within the VChG technique improves tracing accuracy of polarization orientation. In the region we study, a high AM $0.77$ is given between the synthetic dust polarization orientation and Planck 353GHz dust polarization orientation.

\item The new modification of  VChG technique makes it possible to predict degree of dust polarization, which shows a good correspondence with real 353GHz dust map.

\item We successfully give theoretical description of the local gradient angle's statistics.

\item We have demonstrated how to incorporate noise information into VChG analysis.

\end{enumerate}

\section{Acknowledgements}
ZL thanks Ka Ho Yuen for the tutorials of using the sub-block averaging code, sharing his knowledge of data analysis and practical advises during ZL stay at University of Wisconsin, Madison. ZL also thanks Yue Hu for the help in data handling. AL has been supported by NSF AST 1816234 and NASA TCAN AAG1967 grants.
Some of the results in this paper have been derived using the HEALPix \citep{Gorski2005ApJ...622..759G} package. This publication utilizes data from Galactic ALFA HI (GALFA HI) survey data set obtained with the Arecibo L-band Feed Array (ALFA) on the Arecibo 305m telescope. The Arecibo Observatory is operated by SRI International under a cooperative agreement with the National Science Foundation (AST-1100968), and in alliance with Ana G. Méndez-Universidad Metropolitana, and the Universities Space Research Association. The GALFA HI surveys have been funded by the NSF through grants to Columbia University, the University of Wisconsin, and the University of California.




\bibliographystyle{mnras}
\bibliography{main} 





\bsp	
\label{lastpage}
\end{document}